\def\selectedoptions{}

\documentclass[
   \selectedoptions
  ]
  {aipproc}

\layoutstyle{8x11single}
\SetInternalRegister\hbadness{8000} 

\newcommand\doingARLO[2][]{%
  \ifx\mmref\undefined #1\else #2\fi
}

\begin{document}

\title {Ultra-peripheral collisions of relativistic heavy ions}


\author{S. Klein for the STAR Collaboration}{
  address={70-319 LBNL, Berkeley, CA, 94720 USA},
  email={srklein@lbl.gov}, thanks={This work was commissioned by the
  AIP} }

\copyrightyear{2001}

\centerline{Presented at INPC 2001, July 30-Aug. 3, 2001, Berkeley, CA.}

\begin{abstract}

We report the first observation of exclusive $\rho$ production in
ultra-peripheral collisions at RHIC.  The $\rho$ are produced
electromagnetically at large impact parameters where no hadronic
interactions occur.  The produced $\rho$ have a small perpendicular
momentum, consistent with production that is coherent on both the
photon emitting and scattering nuclei.  We observe both exclusive
$\rho$ production, and $\rho$ production accompanied by
electromagnetic dissociation of both nuclei.  We discuss models of
vector meson production and the correlation with nuclear breakup.  We
also observe $e^+e^-$ pair production in these ultra-peripheral
collisions.

\end{abstract}

\date{\today}

\maketitle

\section{Introduction}

Vector mesons can be produced by photonuclear interactions in
ultra-peripheral heavy ion collisions (UPC).  The electromagnetic
interactions occur at impact parameters $b$ larger than twice the
nuclear radius $R_A$, where no hadronic interactions can
occur\cite{baurrev}.  In these UPCs, the electromagnetic field of one
nucleus acts as almost-real photon field, following the
Weizs\"acker-Williams approach.  These photons can fluctuate into a
quark-antiquark pair, which can scatter elastically from the other
nucleus, emerging as a real vector meson.  The photons can also
fluctuate into virtual $\pi^+\pi^-$ pairs, with one of the pions
scattering from the other nucleus and emerging as a real pion pair.

Purely electromagnetic interactions also occur.  At energy scales
above\ \ $\hbar/R_A$, for most purposes, these may be described as
two-photon interactions, although 3 (or more) photon interactions are
also possible\cite{three}.  These reactions can produce $e^+e^-$,
$\mu^+\mu^-$ and $\tau^+\tau^-$ pairs, scalar or tensor mesons, and
meson pairs\cite{background}.  Since the photons couple to charge, the
production rate is a sensitive test of the internal charge content of
mesons, and can be used to rule out glueball candidates.

UPCs occur at moderate impact parameters, $2R_A < b <
\gamma\ \hbar c /M_V$ where $M_V$ is  the vector meson mass
and $\gamma$ the (lab frame) Lorentz boost.  Because the
electromagnetic fields are so strong, additional photonuclear
interactions may accompany the vector meson production.  In addition
to the $\rho$ producing photon, the nuclei may exchange one or more
additional photons which may excite the target nuclei into a giant
dipole resonance or higher excitation.  When the excited nuclei decay,
they emit one or more neutrons.

\begin{figure}
\caption{End and side views of a typical $\rho$ candidate event in
the STAR TPC.  The candidate tracks are almost back-to-back radially,
but boosted longitudinally.}
\resizebox{.45\textwidth}{!}
{\includegraphics[height=.5\textheight,angle=180,clip=true]{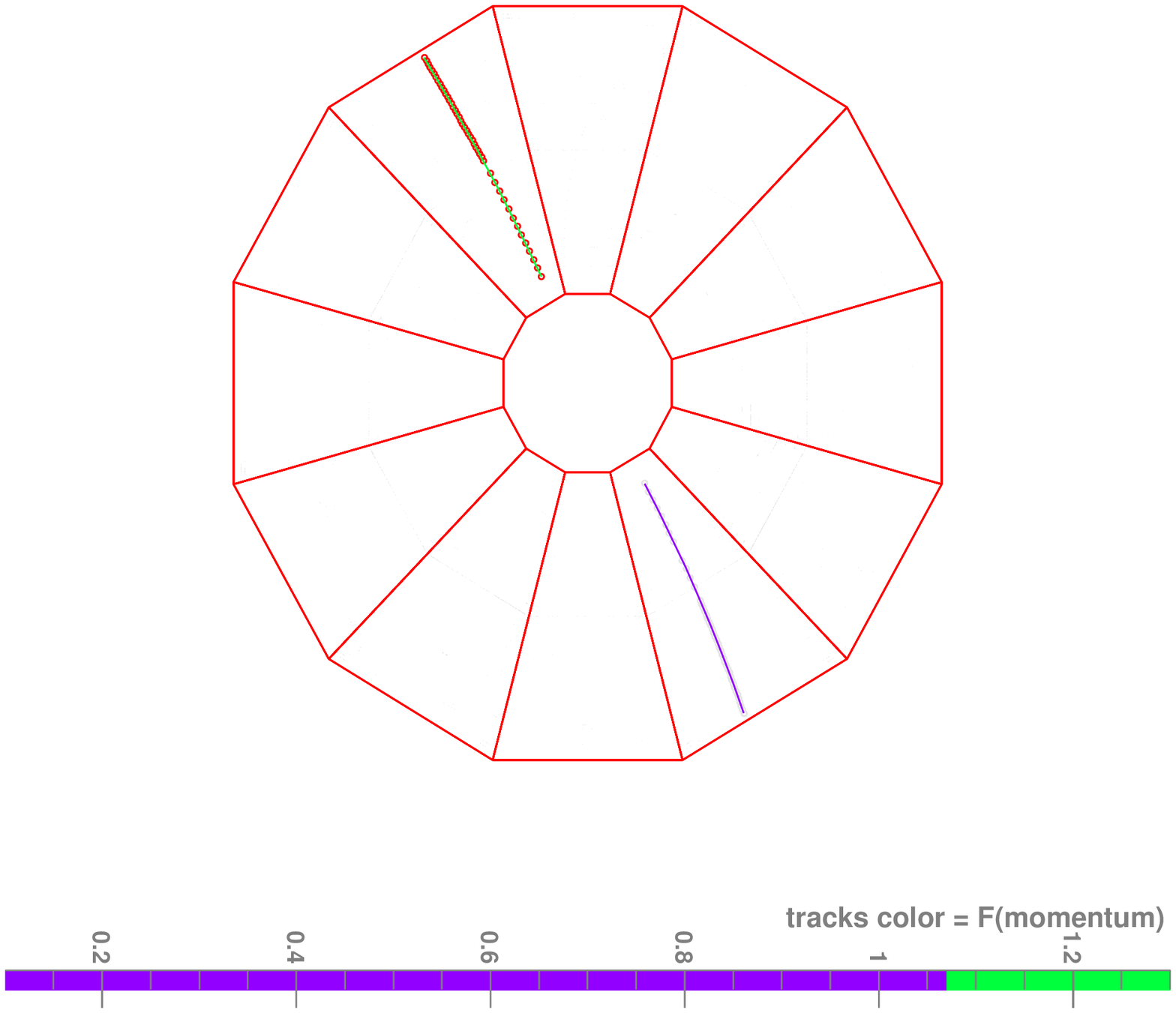}}
\resizebox{.45\textwidth}{!}
{\includegraphics[height=.5\textheight,angle=180,clip=true]{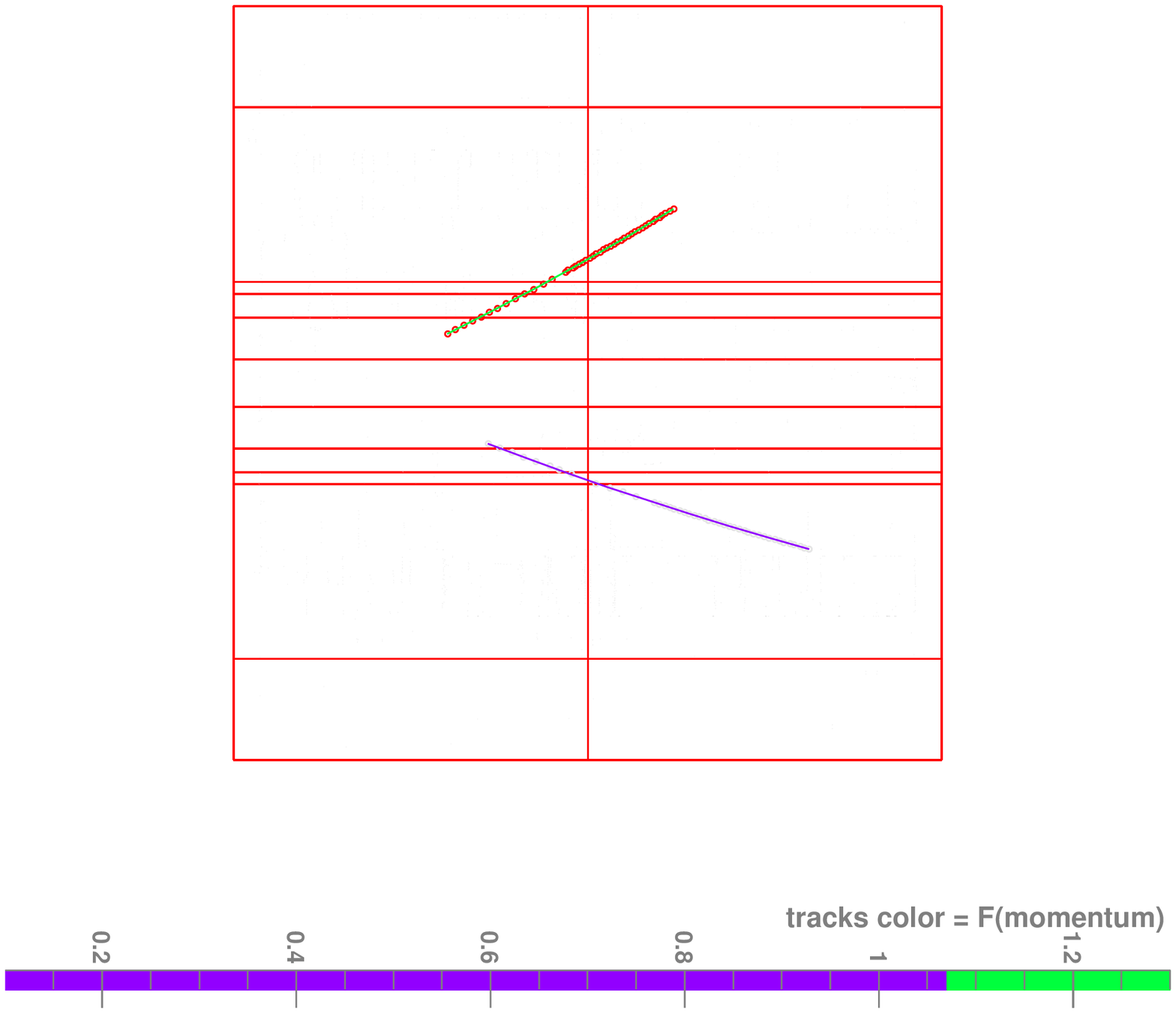}}
\label{fig:event}
\end{figure}

We report the first observation of $\rho$ production in UPCs, both
with and without nuclear excitation.  We also observe $e^+e^-$ pairs
produced in UPCs.  These reactions were observed in $Au + Au$
collisions at a center of mass energy $\sqrt{s_{NN}} = 130$ GeV at the
Relativistic Heavy Ion Collider (RHIC) by the Solenoidal Tracker at
RHIC (STAR) detector.  Figure \ref{fig:event} shows a typical event:
two charged particles are visible in an otherwise empty detector.  The
particles are roughly back-to-back, showing that the pair has a small
transverse momentum $p_T$.

\section{Rates}

The rates for exclusive $\rho$ production may be calculated using data
on $\gamma p\rightarrow \rho p$ collected from HERA and fixed target
experiments.  The photon flux is given by the the
Weizs\"acker-Williams virtual photon method.  The cross section to
produce a vector meson $V$, $\sigma(\gamma A\rightarrow VA)$, is
determined by a Glauber calculation using the $\gamma p\rightarrow Vp$
data as input.  The calculated $\sigma(\gamma A\rightarrow VA)$ cross
sections\cite{vmrates} agree with data from lower energy fixed target
experiments.  For less-well studied mesons, the measured vector meson
rates may be used to determine the meson-nucleon interaction cross
sections.

The production cross section at a given photon energy $k$ (in the lab
frame) maps into the rapidity $\eta$ of the final state vector meson:
$y =1/2 \ln{(2k/M_V)}$.  A photon with a lab-frame energy $k=M_V/2$ is
Lorentz boosted to an energy $\gamma M_V$ in the target frame, so the
$\gamma p$ center of mass energies are comparable to those reached at
Fermilab fixed-target experiments, somewhat below those reached at
HERA.

The total cross section is the integral over $k$ of the photon flux
times $\sigma(\gamma A\rightarrow VA)$.  For gold collisions at
$\sqrt{s_{NN}}=130$ GeV the $\rho$ production cross section is
expected to be about 400 mb, 5\% of the total hadronic cross section.

A $\pi\pi$ final state may be produced directly or through the $\rho$.
The amplitudes for $\rho$ production, $A$, and direct $\pi\pi$
production, $B$, interfere and\cite{soding}
\begin{equation}
{d\sigma\over dM_{\pi\pi}} = \bigg| 
{A\sqrt{M_{\pi\pi}M_\rho \Gamma_\rho} \over M_{\pi\pi}^2 - M_\rho^2
+iM_\rho\Gamma_\rho} + B \bigg|^2
\label{eq:rhosigma}
\end{equation}
where the $\rho$ width is corrected for the increasing phase space as
$m_{\pi\pi}$ increases.  The $\rho$ component undergoes a 180$^o$
phase shift at $M_\rho$, so the interference skews the $\rho$ peak
shape, enhancing production for $M_{\pi\pi} < M_\rho$ and suppressing
the spectrum for $M_{\pi\pi} > M_\rho$.

These UPCs are characterized by final states with small $p_T$ because
of the coherent coupling.  The meson $p_T$ comes from the photon, $p_T
(\gamma) \sim \hbar/b$ and the $p_T$ acquired in the scattering, about
$\hbar/R_A$, added in quadrature.  Vector meson production can occur
at either of the nuclei; because the strong force has a short range,
the production must be inside or very near the nuclei.  Because it is
impossible to determine which nucleus was the source, the amplitudes
from the two production sites (ions) interfere, reducing the number of
$\rho$ with $p_T < \hbar/b$\cite{interfere}.  Because the short-lived
$\rho$ decay before they travel the distance $b$, the interference is
sensitive to the post-decay wave function, and can be used for tests of
quantum mechanics.

Electron-positron pairs may be produced in UPCs by purely
electromagnetic processes.  Although the coupling is large
($Z\alpha\sim 0.6$), in the kinematic range accessible to STAR, higher
order effects are likely to be small, and the cross section is
calculable based on the collision of two almost-real virtual
photons\cite{baurrev}.

Photon exchange can also excite one or both nuclei\cite{baltz}.
Collective excitations, such as the Giant Dipole Resonance, are
possible, along with higher energy incoherent photonuclear
interactions involving hadron production. Calculations indicate that,
in mutual Coulomb dissociation, the two excitations are likely to
occur independently, via two separate photons. The cross section for
mutual coulomb exchange is sizable, around 3.6 barns at $\sqrt{s_{NN}}
= 130$ GeV\cite{mutual}.  This process may be used as a luminosity
monitor.  Mutual excitation also 'tags' events with low impact
parameters.

The impact parameter dependence depends only on the photon flux
variation.  The flux at a given $b$ is given in the
Weizs\"acker-Williams formalism\cite{vidovic}.  Multiple photons from
a single nucleus are emitted independently of each other\cite{gupta}.
As long as there is no interference between final states, the
individual interactions are independent events, with the cross section
for multiple interactions given by an integration of the joint
probabilities over transverse space:
\begin{equation}
\sigma = \int d^2 b P_\rho(b) P_{2EXC}(b) [1-P_{HAD}(b)].
\label{eq:factor}
\end{equation}
where $P_\rho(b)$ is the probability of $\rho$ production and
$P_{2EXC}(b)$ is the probability of mutual nuclear excitation.  There
are some Feynman diagrams for which factorization doesn't apply.  For
example, when photon emission excites the emitting nucleus,
factorization fails.  These processes have been studied for two-photon
UPCs, and the non-factorizing amplitudes were found to be
small\cite{hencken}.  It seems likely that a similar independence
holds for photonuclear interactions.

\section{Data Collection, Triggering and Analysis}

The STAR detector reconstructs charged particles using a 4.2 meter
long time projection chamber (TPC)\cite{TPC} in a 0.25~T solenoidal
magnetic field.  The TPC has an inner radius of 50 cm, and an outer
radius of 2 m.  The charged particle pseudorapidity ($\eta$)
acceptance depends on the position of the interaction.  For the data
discussed here, the interactions were radially within a few cm of the
center of the TPC, but spread longitudinally (in $z$), with $\sigma_z
= 90$ cm.  Tracks originating near $z=0$ (the center of the TPC) were
tracked for $|\eta| < 1.5$.  The momentum resolution was about $\Delta
p/p = 2\% $.  Tracks with $p_T > 100$ MeV/c were reconstructed with
good efficiency.  Particles were identified by their energy loss
($dE/dx$) in the TPC.  For low-multiplicity final states, the $dE/dx$
resolution was about 8\%.

The TPC is surrounded by a cylindrical central trigger barrel (CTB).
For tracks originating near $z=0$, it is sensitive to $|\eta| < 1.0$.
This barrel consists of 240 scintillator slats, each covering $\Delta
y = 0.5$ by $\Delta\phi = \pi/30$.

Two zero degree calorimeters (ZDCs) at $z=\pm 18.25$ meters from the
interaction point detect neutrons from nuclear breakup.  These
calorimeters have $>99\%$ acceptance for single neutrons from nuclear
breakup\cite{zdc}.

The initial trigger decision uses lookup tables and field programmable
gate arrays to initiate TPC readout about 1.5 $\mu$s after the
collision.  TPC readout took 10 msec, allowing for up to 100
events/sec to be read out (at 100\% deadtime).  A higher level filter
uses on-line track reconstruction by a small processor farm \cite{L3}
to remove unwanted events with vertices outside the beam interaction
region.

\section{Exclusive $\rho$ production}

Two separate triggers were used to study $\rho$ production.  The first
('topology') trigger selected events with a small number of tracks
detected by the CTB.  It divided the CTB into 4 azimuthal quadrants,
and required a hit in the opposing 'North' and 'South' quadrants.
Events with hits in the top or bottom quadrants were vetoed as
probable cosmic rays.  We took 7 hours of data with this trigger,
recording about 30,000 events on tape.  The majority of the triggers
were due to cosmic rays, beam-gas events, and debris from upstream
interactions.

\begin{figure}
\caption{(a) The $p_T$ spectrum of topology triggered 2-track
events.  (b) The $m_{\pi\pi}$ spectrum of 2-track events with $p_T <
100$ MeV/c.  The points are oppositely charged pairs, while the
histograms are the like-sign background, scaled up by a factor of
2.1.}
\resizebox{.45\textwidth}{!}{\includegraphics[height=.5\textheight]{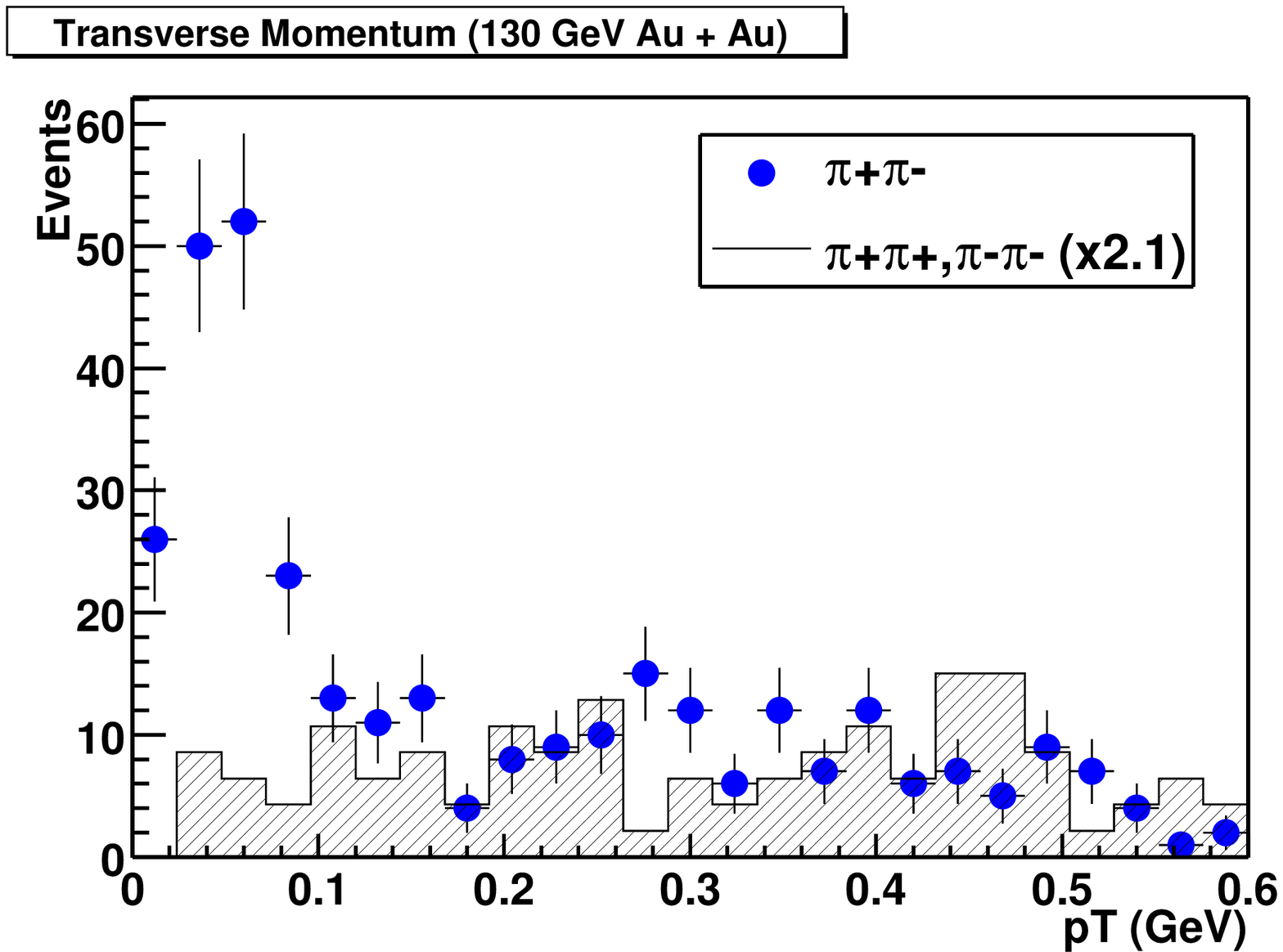}}
\resizebox{.45\textwidth}{!}{\includegraphics[height=.5\textheight]{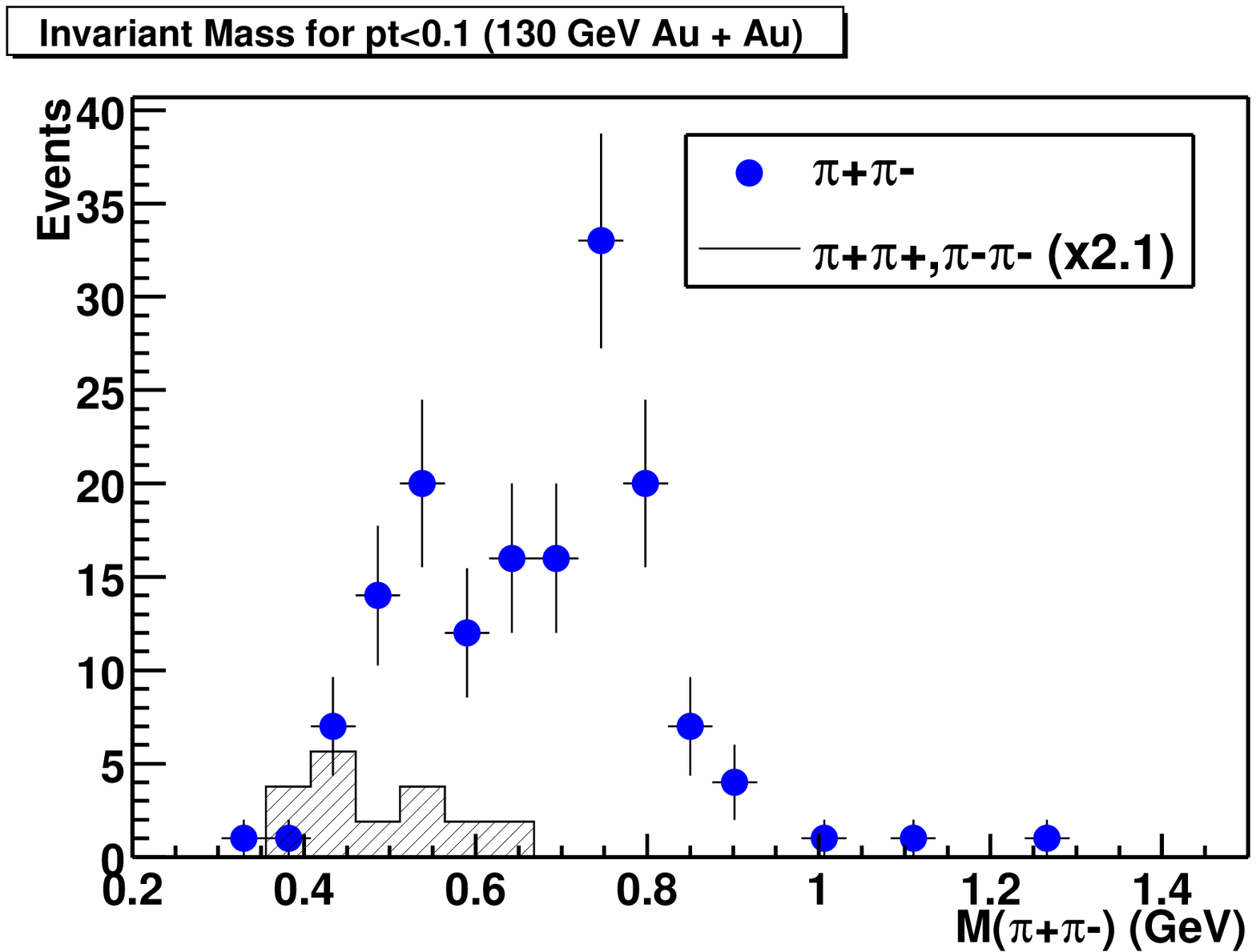}}
\label{fig:pctrig}
\end{figure}

Our analysis selected events with a vertex with exactly 2 tracks
within 2 cm of the centerline of the TPC, and with $|z|< 2$ m.  The
track $dE/dx$ were required to be consistent with pions.  Electrons
and pions can be separated for $p < 140$ MeV/c.  At higher momenta the
two $dE/dx$ bands overlapped.  Finally, to reject the remaining cosmic
rays, we require the events to have a 3-dimensional opening angle
greater than 3 radians (i.e. they must not be perfectly back-to-back).

Fig. \ref{fig:pctrig}(a) shows the pair $p_T$ spectrum of the selected
unlike-sign (dots - net charge 0) and like-sign pairs (histograms).
The unlike pairs are strongly peaked for $p_T<100$ MeV/c.  This is
consistent with production that is coherent with both nuclei; events
with $p_T < 100$ MeV/c are considered our signal.  The like-sign pairs
have no such enhancement, and serve as a background sample.  The
like-sign pairs have been normalized to match the unlike-sign in the
signal-free region 250 MeV $< p_T < $ 500 MeV; this entailed scaling
them up by a factor of 2.1.

Fig. \ref{fig:pctrig}(b) shows the invariant mass of the pairs with
$p_T < 100$ MeV/c.  The points are the unlike-sign events, while the
hatched histogram are the scaled like-sign pairs.  The like-sign pairs
are concentrated at relatively low masses, while the net charge 0
pairs are peaked around the $\rho$ mass.  The asymmetric peak is well
fit by Eq. (1), with $|B/A|$ similar to that found by the ZEUS
collaboration for $\gamma p\rightarrow \rho
p$\cite{ZEUS}\cite{parkcity}.  However, some of the events in the
low-mass shoulder are likely due to $e^+e^-$ pairs; subtracting this
background will decrease $|B/A|$. There is no evidence for any
neutrons in the ZDC data from these events.

\section{$\rho^0$ with nuclear excitation.}

Data on $\rho$ production accompanied by mutual nuclear excitation
were collected with the minimum bias trigger.  This trigger required
that signals from one or more neutrons be detected in each ZDC.

\begin{figure}
\caption{(a) The $p_T$ spectrum for minimum bias 2-track
events.  (b) The $m_{\pi\pi}$ spectrum of 2-track events with $p_T <
100$ MeV/c.  The points are oppositely charged pairs, while the
histograms are the like-sign background, scaled up by a factor of
2.3.}
\resizebox{.45\textwidth}{!}{\includegraphics[height=.5\textheight]{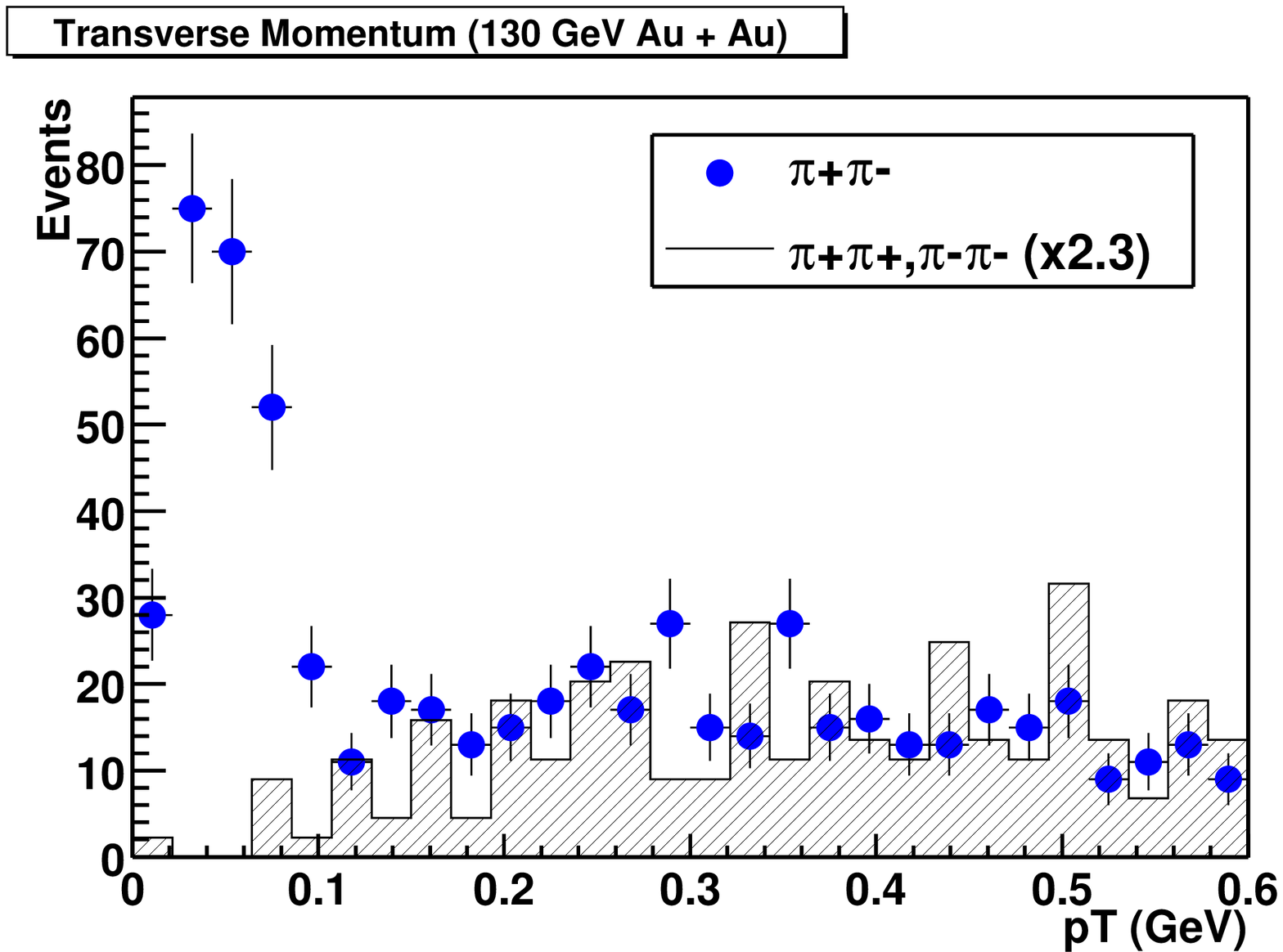}}
\resizebox{.45\textwidth}{!}{\includegraphics[height=.5\textheight]{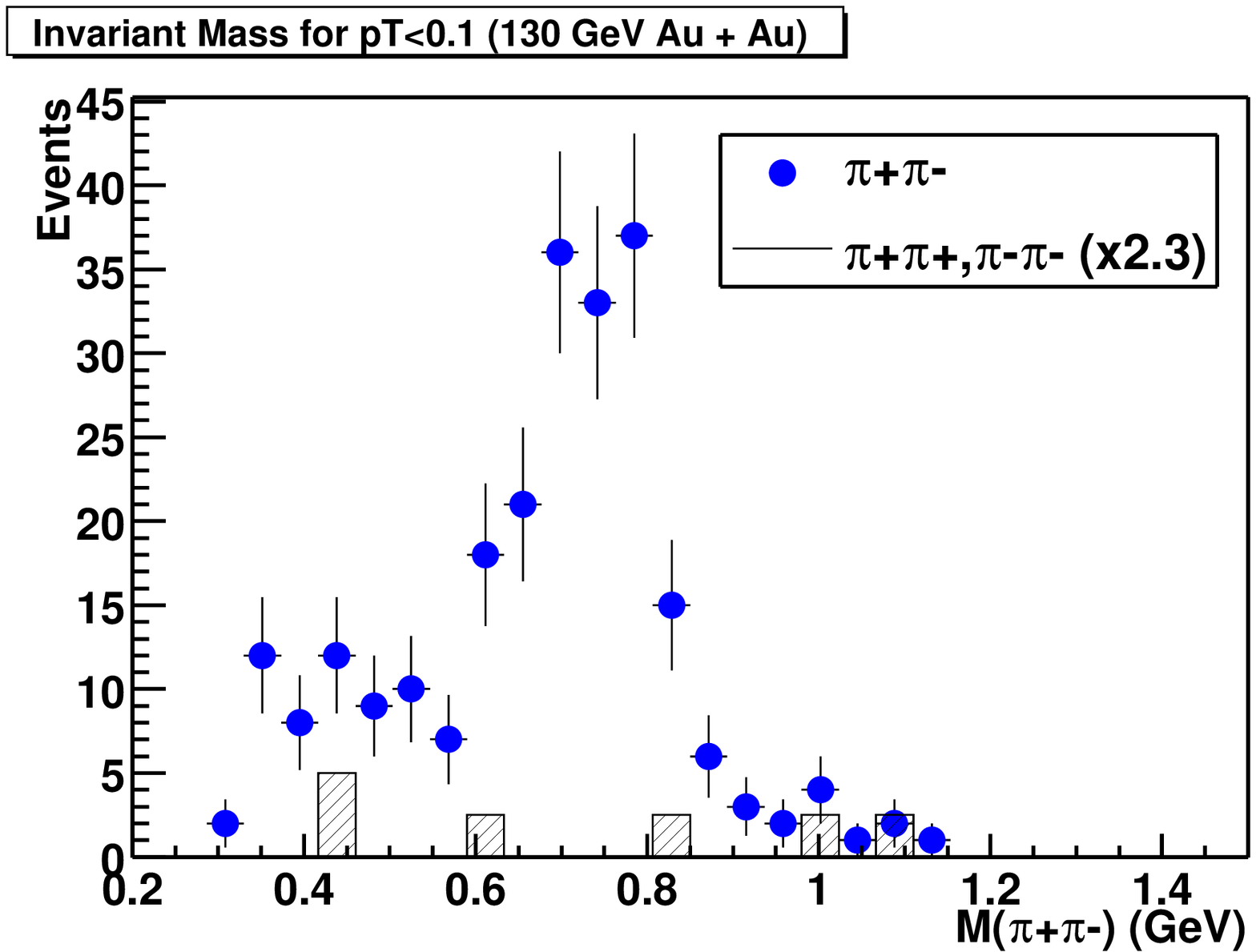}}
\label{fig:mbtrig}
\end{figure}

The event selection for this analysis was the same as for the
topologically triggered sample.  Fig. \ref{fig:mbtrig}a shows the
$p_T$ distribution of the track pairs, and Fig. \ref{fig:mbtrig}b
shows the invariant masses of the $\pi\pi$ combinations with $p_T <
100$ MeV/c.  The overall background is slightly lower than for the
topological trigger, but the composition is quite different.  Here,
the like-sign pair background was scaled up by a factor of 2.3 to
match the like-sign pairs in the region 250 MeV/c $ < p_T < $ 500
MeV/c.  Compared to the topologically triggered sample, the
backgrounds from beam gas events and cosmic rays are greatly reduced,
but the contribution due to grazing nuclear collisions is larger.
Despite these differences, the $\rho$ mass peak has a similar shape
and $|B/A|$ as the topologically triggered data.

\section{Electron-Positron pair production}

\begin{figure}
\caption{(a) The $dE/dx$ of tracks in the 2-track, minimum bias
sample.  The points are from all tracks, while the triangles are from
events where both particles are identified as electrons.  (b) The
$p_T$ spectrum of events where both particles are identified as
electrons.}
\resizebox{.45\textwidth}{!}
{\includegraphics[height=.5\textheight,angle=270]{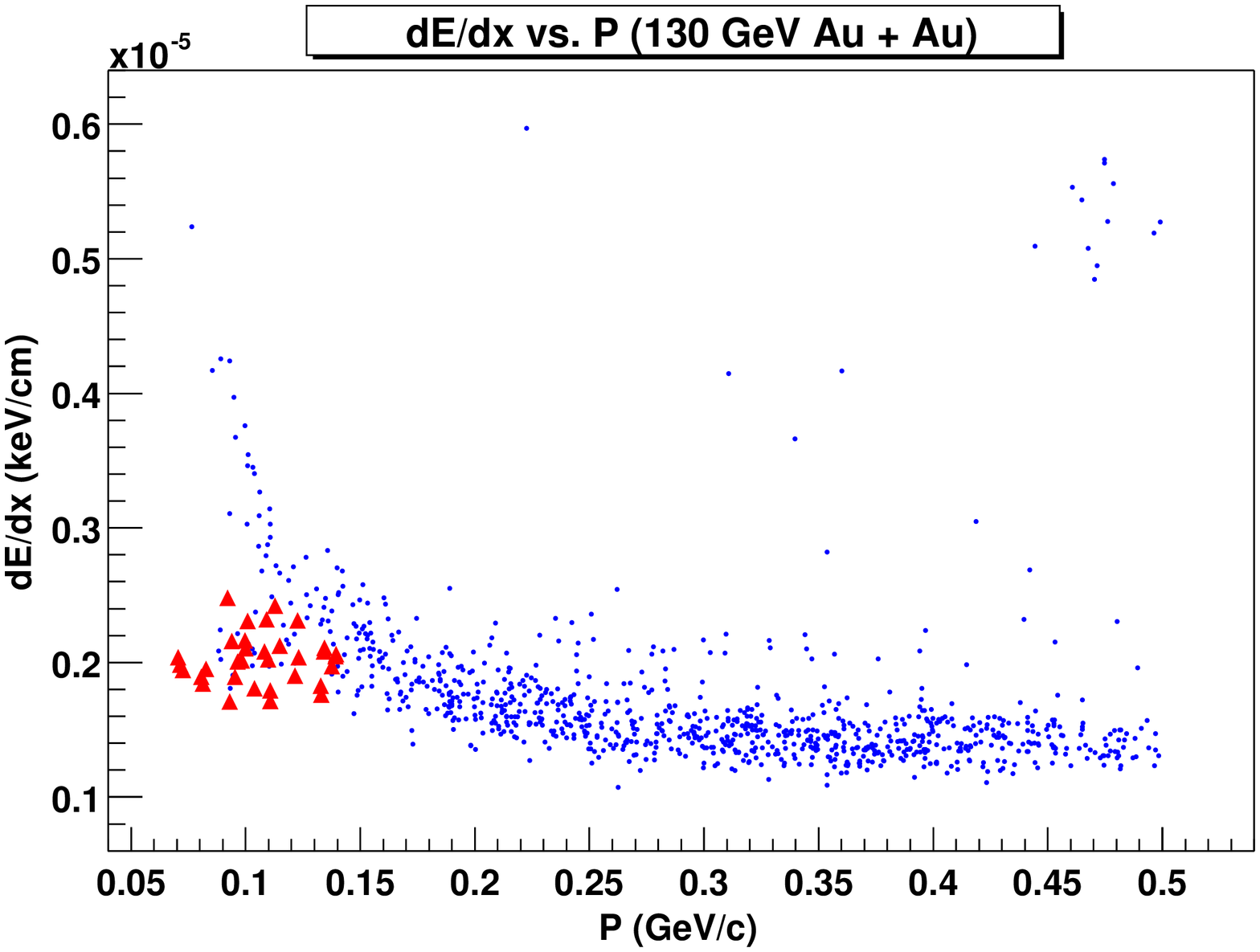}}
\resizebox{.45\textwidth}{!}
{\includegraphics[height=.5\textheight,angle=270]{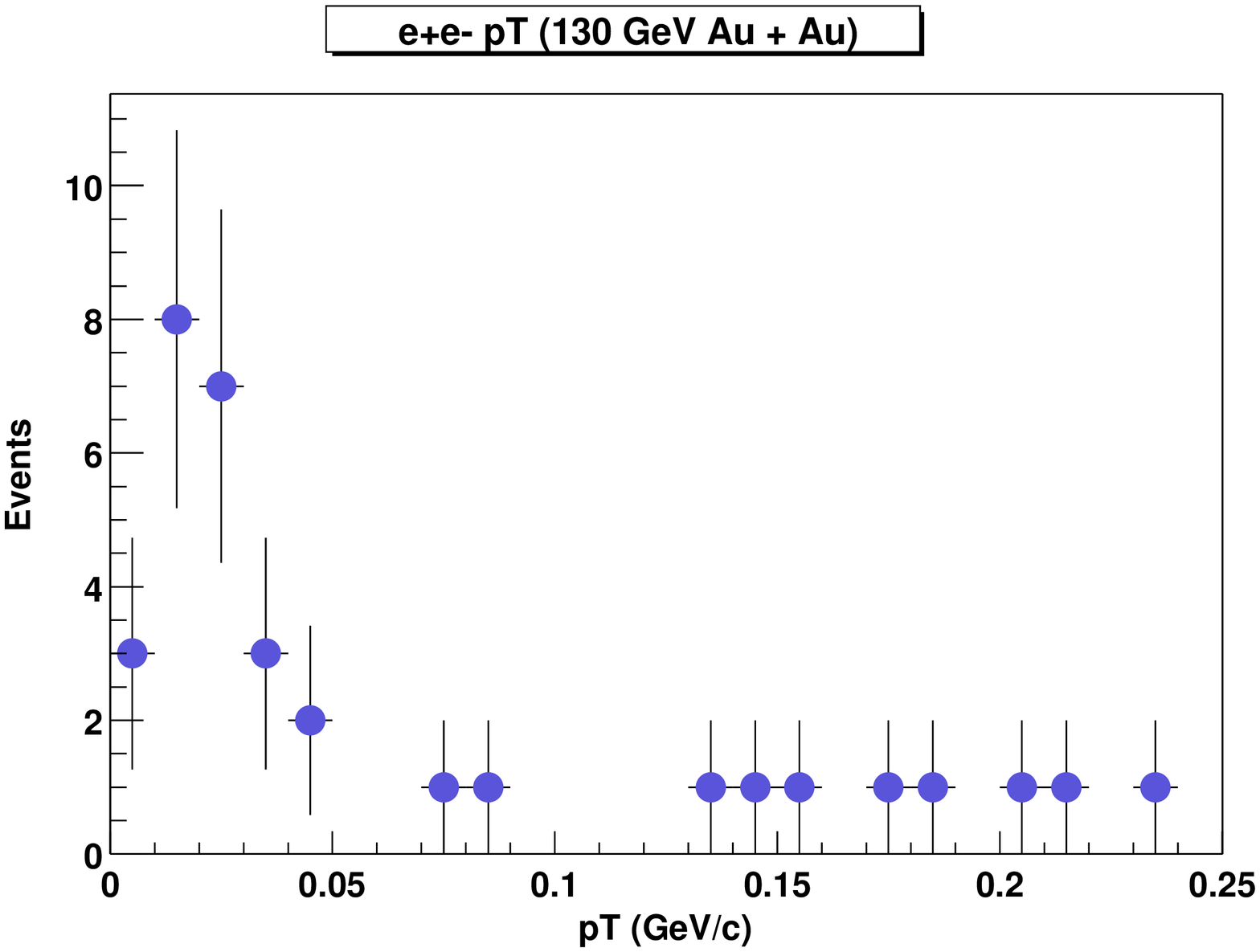}}
\label{fig:electrons}
\end{figure}

This analysis selected events where both particles were identified as
leptons (and hence with $p< 140$ MeV/c).  Leptons from these are shown
by the triangles in
Fig. \ref{fig:electrons}a. Fig. \ref{fig:electrons}b shows the $p_T$
spectrum for events with two identified electrons with $p_T < 140$
MeV/c.  There is a clear peak for pair $p_T < 50$ MeV/c.  The photon
$p_T\sim\hbar c/b$ (considerably less than the\ \ $\hbar c/R_A$
sometimes quoted), so the $p_T$ spectrum is consistent with the
expected photon $p_T$.

\section{Conclusions}

We have observed for the first time the reactions $Au + Au \rightarrow
Au + Au + \rho^0$, $Au + Au \rightarrow Au^* + Au^* + \rho^0$ and $Au
+ Au \rightarrow Au^* + Au^* + e^+e^-$.  The final states have a small
perpendicular momentum, showing their coherent coupling to both
nuclei.

In 2001, RHIC is running at $\sqrt{s_{NN}}= 200 $ GeV and will attempt
to reach full design luminosity, with shorter beam bunches ($\sigma_z
\sim 20$ cm).  STAR has added several new detector elements: a silicon
vertex tracker and forward TPCs, along with an electromagnetic
calorimeter.  In addition, the STAR trigger will be able to take
topological triggers in parallel with central collision triggers.
Together, these improvements should give us several orders of
magnitude more data.  We expect to study additional final states,
including the $J/\psi$ and the $\pi^+\pi^-\pi^+\pi^-$ final states of
the $\rho^*(1450)$ and $\rho^*(1700)$, and to study $\rho$ production
with greater precision and increased angular acceptance.  This will
allow us to study the interference between the two vector meson
production sites.  We will also begin to study two-photon production
of scalar and tensor mesons.


\end{document}